\documentclass[twocolumn,showpacs,prb,citeautoscript]{revtex4}
\usepackage{graphicx}
\usepackage{dcolumn}
\usepackage{bm}
\usepackage{amsmath}
\usepackage{amssymb}
\usepackage{epsfig}
\usepackage{times}

\begin{document}

\title{The proximity effect in normal metal - multiband superconductor
hybrid structures}
\author{A.\ Brinkman}
\author{A.A.\ Golubov}
\affiliation{MESA+ Research Institute and Faculty of Science and Technology, University
of Twente, P.O. Box 217, 7500 AE Enschede, The Netherlands}
\author{M.Yu.\ Kupriyanov}
\affiliation{Institute of Nuclear Physics, Moscow State University, 119992 Moscow, Russia}
\date{\today }

\begin{abstract}
A theory of the proximity effect in normal metal - multiband
superconductor hybrid structures is formulated within the
quasiclassical Green's function formalism. It is shown that the
existence of multiple superconducting bands manifests itself as
the occurence of additional peaks in the density of states in the
structure. The interplay between the proximity effect and the
interband coupling influences the magnitudes of the gaps in a
superconductor in a non-trivial way. The developed theory is
applied to the calculation of supercurrent in multiband
Superconductor - Normal metal - Superconductor Josephson junctions
with low-transparent interfaces, and the results are compared with
the predictions for multiband tunnel junctions.
\end{abstract}

\pacs{74.20.-z, 74.45.+c, 74.70.Ad} \maketitle

The proximity effect is the phenomenon that a superconducting order
parameter can penetrate from a superconductor (S) into a normal metal (N),
or another superconductor (S') with a critical temperature $%
T_{cS^{\prime}}<T_{cS}$, over a distance of the order of the
coherence length, inducing a minigap in N or S. This phenomenon is
well understood, both in terms of Andreev reflections as well as
in terms of microscopic Green's functions \cite{Cooper, Werth,
deGennes, McMillan, KL, Golubov, Fominov, Beenakker, Belzig}.

It is not known, however, how the proximity effect will manifest itself when
multiple pairing potentials are present in the superconductor. This question
has become relevant now that multiband superconductors are coming into
practical use. The most clear example of a multiband superconductor is MgB$%
_2 $, for which the experimental and theoretical evidence for the
coexistence of two gaps is overwhelming \cite{issue}. The multiband nature
of the superconductivity in MgB$_2$ is theoretically well explained \cite%
{Liu} by the qualitative difference between different sheets of the Fermi
surface, together with the large disparity of the electron-phonon
interaction. Therefore, in this paper, the question is addressed how the
multiband nature influences the proximity effect. E.g. what will the density
of states look like in a SN bilayer, where S is a two-band superconductor?

Josephson and quasiparticle tunneling in hybrid structures containing
multiband superconductors have been investigated theoretically in Ref.~ %
\onlinecite{Brinkman} and applied to the calculation of the total
Josephson current in a SIS two-band Josephson tunnel junction. For
all-MgB$_2$ devices, high-quality tunnel barriers are not
available yet, and realizing SNS structures is an attractive
alternative, of which first systems have been realized already
\cite{SNS}. In this paper, the theory of the multiband proximity
effect is applied to the calculation of Josephson current in SNS
structures having two-band S electrodes. The practically
interesting SINIS case is considered, where a non-ideal interface
transparency is taken into account. Predictions are made for
Josephson devices based on MgB$_2$ and compared with those for
MgB$_2$-based tunnel junctions.

In this paper, we will use the quasiclassical Green's function
formalism in order to describe electrical transport in SS' hybrid
structures, where S' is a single-band superconductor while S is a
multiband superconductor. We will restrict ourselves to the limit
of diffusive transport, which is justified if $l_{S,S^{\prime
}}\ll \xi _{S,S^{\prime }}$, where $l_{S,S^{\prime }}$ and $\xi
_{S,S^{\prime }}$ are the electric mean free path and coherence
length of the S and S' materials respectively. In the dirty limit,
the Green's functions in the S' metal are given by the standard
Usadel equations \cite{Usadel}. In the S metal in the regime of
vanishing interband scattering, as is the case for MgB$_{2}$
\cite{cleandirty}, the Usadel
equations take the following form \cite{Koshelev}%
\begin{eqnarray}
\frac{D_{S}^{\alpha }}{2\omega G_{S}^{\alpha }}\left[
(G_{S}^{\alpha })^{2}\Phi _{S}^{\prime \alpha  }\right] ^{\prime
}-\Phi _{S}^{\alpha }=-\Delta
_{\alpha },  \label{Usadel1} \\
\Delta _{\alpha }=2\pi T\sum_{\beta ,\omega \geq 0}\hat{\Lambda}_{\alpha
\beta }\frac{G_{S}^{\beta }\Phi _{S}^{\beta }}{\omega }.  \label{Usadel2}
\end{eqnarray}%
Here, $\alpha $ and $\beta $ are the band indices, e.g. $\alpha,
\beta =1,2$ in the two-band case (later we will use the band
indices $\sigma $ and $\pi $
for MgB$_{2}$ specifically), $\Delta _{\alpha }$ is the pair potential, $%
G_{S}^{\alpha }$ and $\Phi _{S}^{\alpha }$ are Green's functions \cite{KL}, $%
\omega =\pi T(2n+1)$ are Matsubara frequencies, $D_{S}^{\alpha }$
is the diffusion coefficient, and $\hat{\Lambda}_{\alpha \beta }$
is the matrix of effective coupling constants. The prime denotes a
derivative with respect to the coordinate $x$ in the direction
perpendicular to the S-S' interface.

Equations (\ref{Usadel1}) and (\ref{Usadel2}) must in general be
supplemented by boundary conditions. Zaitsev \cite{Zai1} derived boundary
conditions to the quasiclassical Eilenberger equations at the S-S'
boundaries in the clean limit, which were further simplified in Ref.~ %
\onlinecite{KL} in the dirty limit. These boundary conditions have to be
modified when S is a multiband superconductor.

In the limit of small interband scattering a multiband
superconductor may be represented by separate groups of
superconducting electrons which interact with each other only
indirectly, via selfconsistent pair potentials in the bulk.
Therefore, for the derivation of the boundary conditions for the
Usadel equations, one can apply a similar procedure to that used
in Ref.~ \onlinecite{KL} in the single-band case. In the multiband
case, the set of interface parameters, $\gamma ^{\alpha }$ and
$\gamma _{B}^{\alpha }$, describing the proximity effect, should
be introduced for each of the bands.

The first boundary condition relates the current from the S' metal side at
the S-S' interface, $\sigma G^{2}\Phi ^{\prime }$, to that from the S side, $%
\sum_{\alpha }\sigma _{S}^{\alpha }(G_{S}^{\alpha })^{2}(\Phi _{S}^{\alpha
})^{\prime }$. Therefore we have
\begin{equation}
\xi G^{2}\Phi ^{\prime }=\sum_{\alpha }\frac{\xi _{S}^{\alpha }}{\gamma
^{\alpha }}(G_{S}^{\alpha })^{2}(\Phi _{S}^{\alpha })^{\prime },
\label{bound1}
\end{equation}%
with
\begin{equation}
\gamma ^{\alpha }=\frac{\rho _{S}^{\alpha }\xi _{S}^{\alpha }}{\rho \xi }%
,\quad (\xi _{S}^{\alpha })^{2}=\frac{D_{S}^{\alpha }}{2\pi T_{cS}},\quad
\xi ^{2}=\frac{D}{2\pi T_{cS}}\quad
\end{equation}%
Here, $\sigma =1/\rho $ and $\sigma _{S}^{\alpha }=1/\rho
_{S}^{\alpha }$ are
the conductivities of the S' layer and the respective bands of the S metal, $%
D$ is the diffusion constant in S' and $T_{cS}$ is the critical temperature
of S. The ratio between the parameters $\gamma ^{\alpha }$ for the different
bands is mainly determined by the relation between the diffusion constants, $%
D_{S}^{\alpha }$. In the case of MgB$_{2}$, the $\pi $-band is generally
considered to be more dirty than the $\sigma $-band \cite{cleandirty}, i.e. $%
D_{S}^{\pi }\ll D_{S}^{\sigma }$.

The second boundary condition relates the gradient of the Green's
function $\Phi$ near the S-S' interface to its jump at the
interface due to the finite interface resistance,
\begin{equation}
\xi G\Phi ^{\prime }=\sum_{\alpha }\frac{G_{S}^{\alpha }}{\gamma
_{B}^{\alpha }}\left( \Phi _{S}^{\alpha }-\Phi \right) ,  \label{bound2}
\end{equation}%
where $\gamma _{B}^{\alpha }={R_{B}^{\alpha }}/{\rho \xi }$.
$R_{B}^{\alpha } $ are the components of the specific interface
resistance, describing the tunneling of an electron across the
interface into the corresponding conduction band.

In order to obtain the resistances $R_{B}^{\alpha }$, we have to
evaluate the effective junction transparency components. It was
first pointed out by Mazin \cite{Mazin}, that the normal state
conductance $R_{\alpha}^{-1}$, in the limit of a specular barrier
with small transparency, is proportional to the Fermi-surface
average $\left\langle Nv^{2}\right\rangle _{\alpha}$, where $N$ is
the density of states and $v$ the Fermi velocity. In Ref.~ %
\onlinecite{Brinkman} it was further shown that the normal state
resistance component of tunneling into band $\alpha$ of S is given
by the contribution of the electrons in band $\alpha$ to the
squared plasma frequency $\left( \omega _{p}^{\alpha}\right)
^{2}$, which can be obtained from first principle calculations.
For MgB$_{2}$, the ratio $R_{B}^{\sigma }/R_{B}^{\pi }=(\omega
_{p}^{\pi }/\omega _{p}^{\sigma })^{2}$ is $2$ and $100$ for
tunneling in the direction of the $a-b$ plane and $c$-axis
respectively \cite{Brinkman}.

In the case of a SS' bilayer, the Usadel equation (\ref{Usadel1}) needs to
be solved in the S as well as in the S' layer, together with the
self-consistent determination of the pair potentials in S and S', Eq. (\ref%
{Usadel2}). A general numerical method, using $\Theta $-parametrization, $%
\Phi =\omega \mathrm{{tan}\Theta }$ and $G=\mathrm{{cos}\Theta }$, is
described for the single-band case in Ref.~ \onlinecite{Golubov}. Here, we
extended this method by applying the new boundary conditions, Eqs. (\ref%
{bound1}) and (\ref{bound2}). The density of states at energy $E$
can be obtained by applying an analytical continuation, $\omega
=-iE$, to the Usadel equations and the boundary conditions and
solving the numerical scheme in the complex energy plane.

\begin{figure}[tbp]
\includegraphics [scale=1.]{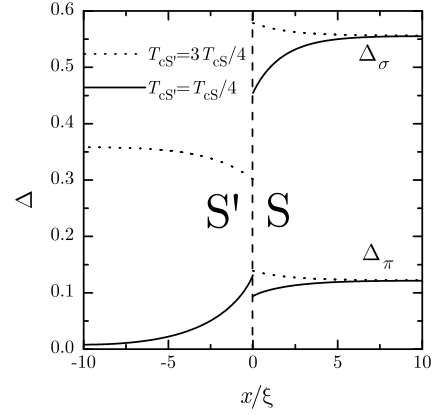}
\caption{Pair-potential as function of position for a SS' bilayer at $%
T=0.5T_{cS}$. The parameters of the bilayer are $\protect\gamma ^{\protect%
\sigma, \protect\pi }=1$, $\protect\gamma _{B}^{\protect\sigma }=2$, $%
\protect\gamma _{B}^{\protect\pi }=1$, $d_{S}/\protect\xi
_{S}=d_{S^{\prime }}/\protect\xi _{S^{\prime }}=10$, and the
coupling constants in the S-layer are chosen as expected
\protect\cite{const} for MgB$_{2}$: $\Lambda _{11}=0.81$, $\Lambda
_{22}=0.278$, $\Lambda _{12}=0.115$, and $\Lambda _{21}=0.091$.}
\label{pairpot}
\end{figure}

The numerically obtained dependence of the pair-potential on position is
presented in Fig. \ref{pairpot} for the example in which the coupling
constants are taken as calculated for MgB$_{2}$ in Ref.~ %
\onlinecite{const}. The parameter values are indicated in the
caption.
For temperatures above $T_{cS^{\prime }}$ (solid lines in Fig. \ref{pairpot}%
), it can be seen that the pair-potential in S' increases towards
the interface, while $\Delta _{\sigma }$ decreases, as expected in
analogy with the proximity effect in the single-band case. The
decrease in $\Delta _{\sigma }$ towards the interface can be
explained by the relatively strong coupling between the $\sigma $
and $\pi $ bands. By decreasing the interband coupling constants
and by decreasing the interface suppression parameters, one can
obtain the opposite regime, in which $\Delta _{\pi }$
increases towards the interface. For relatively large values of $%
T_{cS^{\prime }}$, and for temperatures below $T_{cS^{\prime }}$, we even
predict an increase in $\Delta _{\sigma }$ towards the interface, as
illustrated by the dashed line in Fig. \ref{pairpot}. This effect is further
enhanced when the $\sigma $-band is decoupled from the S'-layer, which is
the case for example when the interface normal is parallel to the
crystallographic $c$-axis of MgB$_{2}$, due to the vanishingly small ratio $%
R_{B}^{\pi }/R_{B}^{\sigma }$ in that case.

\begin{figure}[tbp]
\includegraphics [scale=1.]{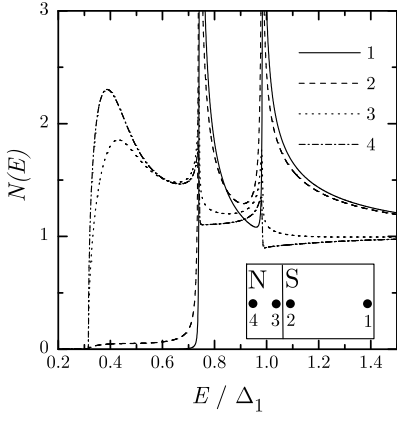}
\caption{Normalized density of states in a proximized SN-bilayer
at several positions in the bilayer ($1-4$, as indicated in the
inset), where S is a two-band
superconductor. The parameters of the bilayer are $\protect\gamma ^{1,2}=0.1$, $\protect\gamma %
_{B}^{1,2}=5$, $d_{S}/\protect\xi _{S}=10$, $d/\protect\xi =1$ and the
coupling constants in the S-layer are chosen as $\Lambda _{11}=0.5$, $%
\Lambda _{22}=0.4$, $\Lambda _{12}=\Lambda _{21}=0.1$.}
\label{prox_dos}
\end{figure}

As an example, in Fig. \ref{prox_dos} the results of a calculation
of the density of states in a SN-bilayer are presented. In the
considered case, the bulk energy gaps in a two-band superconductor
are not too different. As is seen from the figure, the density of
states in the N-layer has three peaks: the lowest energy peak
corresponds to the proximity induced minigap and the two other
peaks correspond to the bulk energy gaps in the two-band
superconductor S. The existence of a minigap is a characteristic
feature of the proximity effect in a SN-bilayer in the dirty
limit, as was studied in detail in the single-band case in Ref.~
\onlinecite{Golubov}. As we can see, the minigap persists in the
two-band case as well and its magnitude depends on the parameters
of the interface, thicknesses of the N and S and the values of the
bulk gaps in the superconductor.

The next step in investigating the influence of multiband superconductivity
on the proximity effect, is to study supercurrents in multiband proximized
structures. We will consider double-barrier structures consisting of two S
electrodes coupled by a normal metal N. As a model system we use a SINIS
double-barrier hybrid structure, since in practical devices interface
potential barriers are always present at the S-N interfaces, either
originating from a Fermi-velocity mismatch, degradation of surface layers,
or artificially deposited oxide barriers.

If the conditions of the dirty limit (electron mean free path
$\ell \ll d,\xi $) are fulfilled in the N interlayer, than the
stationary Josephson effect in the structure can be analyzed in
the framework of the Usadel equations by the method developed in
Refs. ~\onlinecite{KL} and ~\onlinecite{Zai1} for the single-band
case. We assume that the interface transparencies are small enough
such that the condition $1+\gamma _{B1,2}^{\alpha }\gg
\gamma_{1,2} ^{\alpha }$ holds at both NS interfaces (here and
below we drop the subscript $S$). In this case, the suppression of
superconductivity in the S layers is weak and the Green's
functions in the electrodes near the interfaces, $G_{1,2}^{\alpha
}$ and $\Phi _{1,2}^{\alpha }$, are equal to their bulk values. To
calculate the supercurrent, it is sufficient to consider Eq.
(\ref{bound2}) at the two interfaces, giving
\begin{equation}
\xi G\Phi ^{\prime }=\sum_{\alpha }\frac{G_{1,2}^{\alpha }}{\gamma
_{B1,2}^{\alpha }}\left( \pm \Phi _{1,2}^{\alpha }\mp \Phi \right)
,\quad x=\pm \frac{d}{2}. \label{bcd1}
\end{equation}

For simplicity, we will consider symmetric junctions where
$G_{1,2}^{\alpha }\equiv G_{S}^{\alpha }$ and $\gamma
_{B1,2}^{\alpha }\equiv \gamma _{B}^{\alpha }$, and where the
functions $\Phi _{1,2}^{\alpha }$ are related to the phase shift
$\varphi $ across the junction by $\Phi _{1,2}^{\alpha }=\Delta
_{\alpha }\exp \left( \pm i\varphi /2\right)$. Further, we
consider purely normal N-layer with vanishing pair potential
$\Delta =0$ and restrict ourselves to considering the limit of a
small interlayer thickness, $d\ll \xi $.

In the limit $d\ll \xi $, there are two characteristic frequencies
$\Omega _{1,2}$ in the Usadel equations
(\ref{Usadel1},\ref{Usadel2}). At $\omega \lesssim \Omega _{1}=\pi
T_{cS}\frac{\xi }{d}\gg \Omega _{2}=\pi T_{cS}$ we can neglect all
nongradient terms in the Usadel equation. Hence, $\left[
G^{2}\Phi ^{\prime }\right] ^{\prime }=0$, and in the zero approximation on $%
d/\xi $ one obtaines that all $\Phi $ functions are spatially independent
constants, $\Phi =A$. In the next approximation we have
\begin{equation}
\Phi =A+B\frac{x}{\xi }+A\frac{x^{2}\beta ^{2}}{2\xi ^{2}},\quad \beta ^{2}=%
\frac{\omega }{\pi T_{cS}G}.  \label{Solu_1}
\end{equation}%
From the boundary conditions and by taking into account that in our model $%
\Phi _{1,2}^{\alpha }=\Delta _{\alpha }\exp \left( \pm i\varphi /2\right) $,
we finally will have
\begin{eqnarray}
A &=&\frac{\widetilde{\Delta }\eta }{\widetilde{G}},\quad B=\frac{i%
\widetilde{\Delta }}{G}\frac{d}{\xi }\sin (\varphi /2),  \label{AB} \\
G &=&\frac{\omega }{\sqrt{\omega ^{2}+A^{2}}}=\frac{\omega \widetilde{G}}{%
\sqrt{\omega ^{2}\widetilde{G}^{2}+\widetilde{\Delta }^{2}\eta ^{2}}},
\label{G}
\end{eqnarray}%
where $\gamma _{BM}^{\alpha }=\gamma _{B}^{\alpha }d/\xi $, $\eta ^{2}=\cos
^{2}(\varphi /2)$, and
\begin{equation}
\widetilde{G}=\sum_{\alpha }\frac{G_{S}^{\alpha }}{\gamma _{BM}^{\alpha }}+%
\frac{\omega }{2\pi T_{cS}},\quad \widetilde{\Delta }=\sum_{\alpha }\frac{%
G_{S}^{\alpha }\Delta _{\alpha} }{\gamma _{BM}^{\alpha }}.
\end{equation}

\begin{figure}[tbp]
\includegraphics [scale=1.]{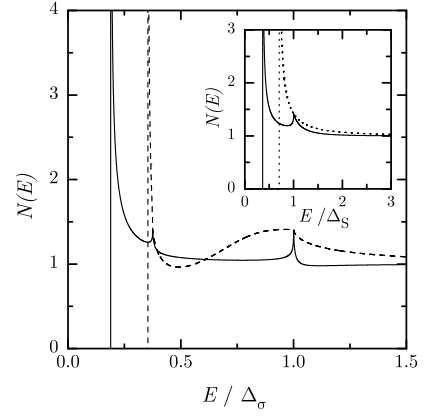} .
\caption{Normalized density of states in the interlayer of a SINIS
double-barrier structure, where S is the two-band superconductor MgB$_{2}$,
and the phase difference over the junction is $\protect\pi /2$. The density
of states is shown for small $\protect\gamma _{BM}$ (dashed line: $\protect%
\gamma _{BM}^{\protect\sigma }=0.2$, $\protect\gamma _{BM}^{\protect\pi %
}=0.1 $) and large $\protect\gamma _{BM}$ (solid line: $\protect\gamma %
_{BM}^{\protect\sigma }=20$, $\protect\gamma _{BM}^{\protect\pi }=10$). For
comparison, the inset shows the density of states in the interlayer of a
SINIS junction with single-band superconductors (solid line: $\protect\gamma %
_{BM}=2$, dashed line $\protect\gamma _{BM}=2.10^{-3}$).}
\label{prox_SNS}
\end{figure}

The density of states $N(E)=\mathrm{{{Re}(G)}}$ in the interlayer of the
double-barrier junction can now be found from an analytical continuation of
Eq. (\ref{G}) to real energies, $\omega =-iE$. The results for the two-band
case are plotted in Fig. \ref{prox_SNS}. The known density of states for a
single-band SINIS junction \cite{PRB} is shown in the inset. For $\gamma
_{BM}\ll 1$, the single-band results show a peak in the density of states at
$\Delta \mathrm{{cos}(\phi /2)}$, while the density of states in the
two-band junction in this regime is predicted to have a peak at a value that
is even lower than $\Delta _{\pi }\mathrm{{cos}(\phi /2)}$. For larger
values of $\gamma _{BM}$, the density of states shows three peaks: at the
minigap and at $\Delta _{\pi }$ and $\Delta _{\sigma }$, in analogy with the
two peaks in the density of states of a single-band SINIS junction.

Substituting Eq. (\ref{AB}) into the supercurrent expression,
\begin{equation}
I=\sigma 2\pi T\mathrm{{Im}\sum_{\omega \geq 0}\frac{1}{\omega ^{2}}%
G^{2}\Phi ^{\ast }\Phi ^{\prime },}  \label{currentdef}
\end{equation}%
and taking $\Phi ^{\ast }\Phi ^{\prime }$ in lowest order equal to $A^{\ast
}B$, we obtain
\begin{equation}
I=\frac{\pi T}{\xi \rho }\sum_{\omega \geq 0}\frac{
d\widetilde{\Delta }^2\sin (\varphi )}{\omega \xi \sqrt{\omega ^{2}\widetilde{G}^{2}+\widetilde{\Delta }^{2}\eta ^{2}}}%
.  \label{curr_fin}
\end{equation}%
A generalization to take boundary asymmetry and a finite $\Delta $ in the
interlayer into account can be made straightforwardly.

In the two-band case in the limit $\gamma _{BM}^{\sigma }\rightarrow \infty$%
, which is for example the case for tunneling in the MgB$_2$ $c$-axis
direction, the normal metal is only proximized by the $\pi$-gap of electrode
S and Eq. (\ref{curr_fin}) gives
\begin{equation}
I=\frac{\pi T}{\xi \rho \gamma _{B}^{\pi }}\sum_{\omega \geq 0}\frac{%
G_{S}^{\pi }\Delta _{\pi }^{2}\sin (\varphi )}{\omega \sqrt{\left[ \omega +%
\frac{\omega ^{2}\gamma _{BM}^{\pi }}{2\pi T_{cS}G_{S}^{\pi }}\right]
^{2}+\Delta _{\pi }^{2}\cos ^{2}(\frac{\varphi }{2})}},  \label{curr_1gap}
\end{equation}%
which has been previously obtained \cite{KL,Kupr} for SINIS junctions with
single-band superconductivity in S. It also follows from Eqs. (\ref{curr_fin}%
) and (\ref{curr_1gap}) that the critical current of a single-band
SINIS junction always exceeds the critical current of a two-band
SINIS junction with a vanishing second gap, $\Delta _{\pi }=0$.

\begin{figure}[tbp]
\includegraphics [scale=1.]{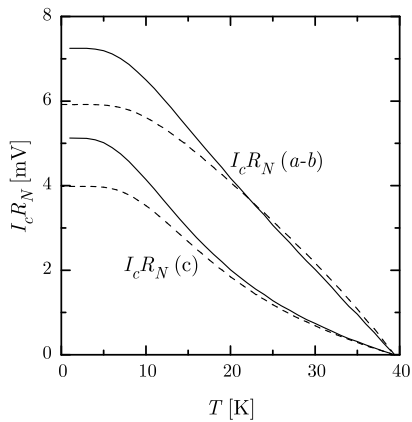}.
\caption{$I_{c}R_{N}$ for double-barrier MgB$_{2}$ SINIS junctions
in the regime of $\gamma _{BM}^{\sigma,\pi} \ll 1$ (solid lines),
compared to $I_{c}R_{N}$ for MgB$_{2}$ SIS tunnel
junctions\protect\cite{Brinkman} (dashed lines). The total
$I_{c}R_{N}$ of $a-b$ plane MgB$_{2}$ junctions is an average over
all bands, while $c$-axis junctions only contain a $\protect\pi
$-band contribution.} \label{Ic_SNS}
\end{figure}

The temperature dependence of the critical current can now be calculated for
SINIS Josephson structures for different orientations of the
crystallographical axis with respect to the interface normal. The gap
functions, $\Delta _{\pi ,\sigma }(T)$ and the ratio $\gamma _{B}^{\sigma
}/\gamma _{B}^{\pi }$, follow from band structure calculations \cite%
{Brinkman}. The results are shown in Fig. \ref{Ic_SNS} for
vanishingly small $\gamma _{{BM}}$, and compared to the
calculation results for SIS junctions \cite{Brinkman}. The full
specific interface resistance of a SINIS junction,
$R_{N}=R_{B}^{\sigma }R_{B}^{\pi }/[R_{B}^{\sigma }{+}R_{B}^{\pi
}]$. It is clearly seen that the critical current of SINIS
junctions is larger than in SIS structures, practically in the
whole temperature region, as is the case for single band
superconductors \cite{KL,Kupr}. At low temperatures the
$I_{c}R_{N}$ product can be as large as 5.2 mV when only the $\pi
$-band contributes to the current and close to 7.3 mV when the sum
over different band contributions can be taken into account, as is
the case for tunneling in the direction of the $a-b$ plane. The
negative curvature of $I_cR_N(T)$ is a direct consequence of the
two-band nature of superconductivity and is absent in $I_cR_N(T)$
of single-band SINIS junctions in the regime of small $\gamma
_{BM}$ \cite{Kupr}.

In summary, we have formulated a microscopic theory of the
proximity effect in hybrid structures based on multiband
superconductors in the diffusive limit. We have shown that the
existence of multiple superconducting bands manifests itself in
the proximity effect between a normal metal and a superconductor
as the occurence of additional peaks in the density of states at
the normal metal side. The interplay between the proximity effect
and interband coupling determines the gap magnitudes at the
interfaces. The supercurrent in multiband proximized Josephson
junctions was calculated and compared to known single-band results
and predictions for multiband tunnel junctions.

We acknowledge useful discussions with O.V. Dolgov and A.E. Koshelev and
support from INTAS project 2001-0617. M.Yu.K. acknowledges support from the
Russian Ministry of Industry, Science and Technology.

\end{document}